\documentclass[a4paper,11pt]{article}
\usepackage{pos}

\usepackage{graphicx}
\usepackage{multirow} 
\usepackage{amsmath}
\usepackage{enumerate}
\usepackage{subcaption} 
\usepackage{color} 
\usepackage{wrapfig}
\usepackage[normalem]{ulem}
\usepackage[utf8]{inputenc}
\usepackage[outdir=./]{epstopdf}

\newcommand{\zhaires}{\mbox{ZHA{\scriptsize{${\textrm{IRE}}$}}{S }}}

\title{RDSim: A fast, accurate and flexible framework for the simulation of the radio emission and detection of downgoing air showers.}

\author*[a]{Washington R. de Carvalho Jr.}
\author[a]{Abha Khakurdikar}

\affiliation[a]{IMAPP, Radboud University,\\ Heyendaalseweg 135, Nijmegen, The Netherlands\\ }

\emailAdd{carvajr@gmail.com}
\emailAdd{a.khakurdikar@astro.ru.nl}

\abstract{RDSim is a fast, accurate and flexible framework for the simulation of the radio emission of downgoing air showers and its detection by an arbitrary array, including showers initiated by neutrino interactions or tau-lepton decays. RDSim was build around speed and is based on simple and fast, yet still accurate, toymodel-like approaches. It models the radio emission using a superposition emission model that disentangles the Askaryan and geomagnetic components of the shower radio emission. It uses full \zhaires simulations as an input to estimate the electric field at any position on the ground. A single input simulation can be scaled in energy and rotated in azimuth, taking into account all relevant effects. This makes it possible to simulate a huge number of geometries and energies using just a few \zhaires input simulations. RDSim takes into account the main characteristics of the detector, such as trigger setups, thresholds and antenna patterns. To accommodate arrays that use particle detectors for triggering, such as the Auger RD extension, it also features a second toymodel to estimate the muon density at ground level and perform simple particle trigger simulations. Owing to the large statistics made possible by its speed, it can be used to investigate in detail events with a very low trigger probability and geometrical effects due to the array layout, making it specially suited to be used as a fast and accurate aperture calculator. In case more detailed studies of the radio emission and detector response are desired, RDSim can also be used to sweep the phase-space for the efficient creation of dedicated full simulation sets. This is particularly important in the case of neutrino events, that have extra variables that greatly impact shower characteristics, such as interaction or $\tau$ decay depth as well as the type of interaction and it's fluctuations.}

\FullConference{%
  The 27th European Cosmic Ray Symposium - ECRS 2022 \\
  25-29 July 2022\\
  Nijmegen, The Netherlands}

\begin{document}
\maketitle

\section{Introduction}

RDSim is a framework for the simulation of the radio emission and detection of extensive air showers (EAS) by an arbitrary antenna array. The speed of the simulation set-up makes it possible to investigate larger areas around the detector, events with very low detection probability and examine the impact of asymmetries and border effects of the array. RDSim is specially suited to be used as a fast and accurate aperture calculator.


This work is organized as follows: Section \ref{sec:emission} describes the radio emission model used while on section~\ref{sec:detection} we describe the simple detector response models and the overall structure of RDSim, including a description of the optional particle trigger simulation. Section~\ref{sec:neutrinos} outlines the extra models used in the case of neutrino events, such as sampling of the interaction point and tau-lepton propagation, followed by a discussion on Section~\ref{sec:discussion}.

\section{Radio emission model}
\label{sec:emission}

\begin{figure}
  \begin{center}

    \vspace{-0.8cm}
    \includegraphics[width=0.25\textwidth]{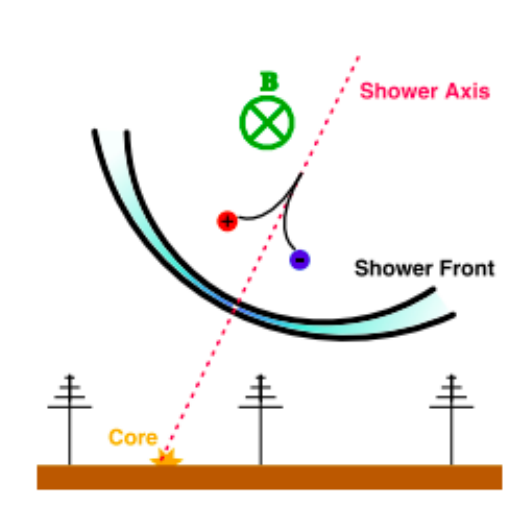}\hspace{2cm}\includegraphics[width=0.25\textwidth]{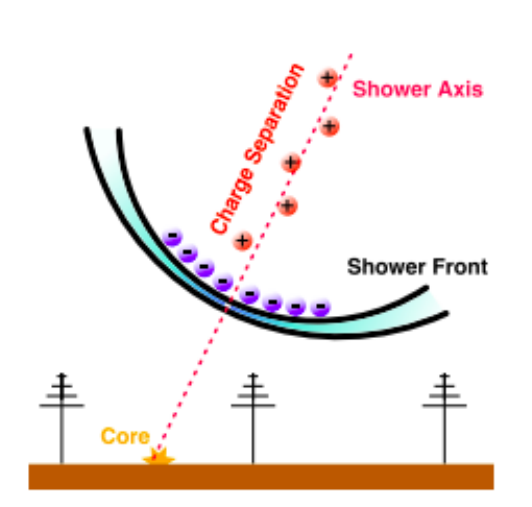}
    
    \includegraphics[width=0.36\linewidth, trim= 0 -2cm  0 0.8cm, clip]{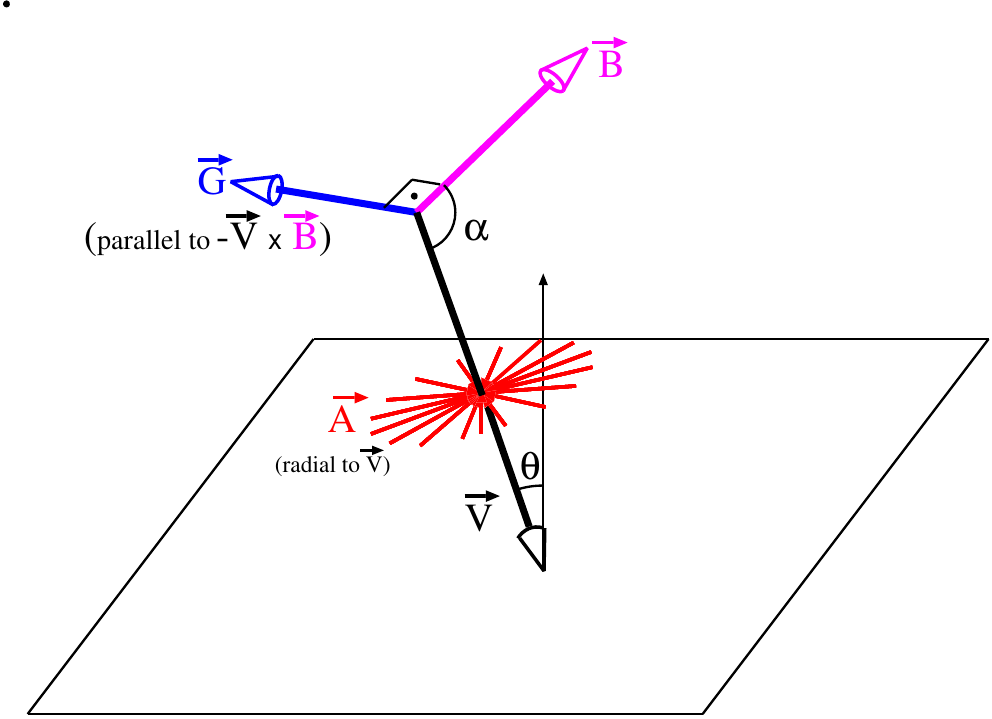}\hspace{1cm}\includegraphics[width=0.41\linewidth]{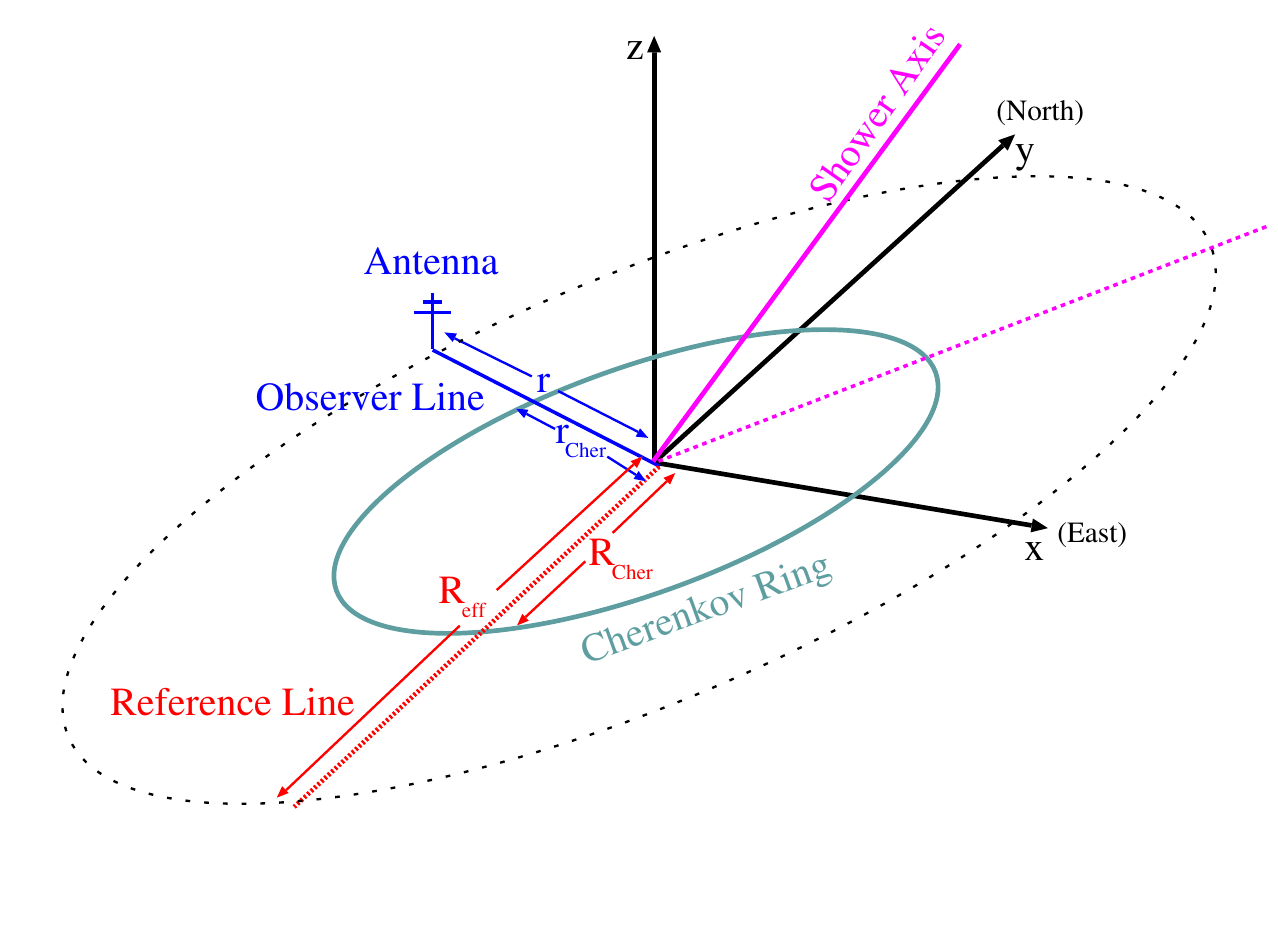}
  \end{center}
  
  \vspace{-1.2cm}
  
  \caption{Top: Geomagnetic emission mechanism (left) and charge excess mechanism (right). Figures extracted from~\cite{schoorlemmer2012tuning}. Bottom left: Theoretical polarization of the Askaryan (red) and geomagnetic (blue) emission mechanisms. The shower axis ($\vec{v}$) is shown in black and the geomagnetic field $\vec{B}$ in magenta. Bottom right: Assumed elliptical symmetry at ground level, described by an ellipse with its major axis aligned with the projection of the shower axis on the ground. Both figures were extracted from \cite{toymodel}. }
  \label{fig:toymodelscheme} 
\end{figure}

  When the air shower progresses in the atmosphere, the electromagnetic component of the shower emits radio emission. The measured radio emission is the composite of different emission mechanisms. The two main contributors to the radio emission are known as the Geomagnetic and Askaryan mechanisms. When the secondary electrons and positrons of the air shower travel through the atmosphere, they are accelerated by the Earth's magnetic field. The accelerated electrons and positrons drift apart in opposite directions (see top left of Fig. \ref{fig:toymodelscheme}). The polarization direction of the resulting current is thus in the direction of the Lorentz force. As the air shower evolves, the number of secondary particles increases, reach maximum and then die out. Macroscopically, these accelerated particles create a time varying current that emits radiation. This is known as the geomagnetic emission mechanism and accounts for \(80-90\%\) of the radio emission of the extensive air shower. It is roughly proportional to the intensity of the Lorentz force and thus to $\sin{\alpha}$, where $\alpha$ is the angle between the shower axis and the geomagnetic field (see bottom left panel of Fig. \ref{fig:toymodelscheme}).



The remaining \(10-20\%\) of the radio emission contribution comes from the Charge-excess mechanism, also known as Askaryan effect. As the charged particles interact with the air molecules, they ionize the medium. As the shower proceeds, the free electrons are then swept along with the shower front (see top right panel of Fig. \ref{fig:toymodelscheme}), creating an excess of electrons in the shower. Macroscopically this can be thought as a dipole that moves along with the shower front due to the charge separation. The radio emission of this mechanism is polarized radially with respect to the shower axis \cite{HUEGE20161} (see also the bottom left panel of Fig. \ref{fig:toymodelscheme}).


The radio emission model in RDSim is based on the superposition of the Geomagnetic and Askaryan emission mechanisms and is an expansion of the model described in \cite{toymodel}. The emission of both mechanisms is roughly linearly polarized, but due to their different polarizations the superposition of these two mechanisms leads to an asymmetric footprint of the observed radiation pattern on ground. The RDSim emission model uses ZHAireS simulations as an input. ZHAireS \cite{zhaires} is an extension of AIRES (AIRshower Extended Simulations) \cite{aires} used to calculate the radio emission of air showers using the ZHS algorithm, which is based on first principles and do not presuppose any emission mechanisms. Instead it calculates the contribution to the vector potential of every single charged particle track in the shower. ZHAireS simulations of just a few antennas along a reference line are performed and RDSim disentangles the two emission components in order to obtain the peak electric field amplitude, for each mechanism separately, as a function of the distance $r$ to the core along the reference line.  The model then assumes an elliptical symmetry at ground level (see bottom right panel of Fig.\ref{fig:toymodelscheme}) for the amplitudes of each mechanism and combines them, along with their respective polarizations, to obtain the net electric field and its polarization at any position on the ground. The model takes into account early-late effects that arise due to the varying distance to the shower as the observer position changes and also features a simple linear scaling of the electric field with shower energy.

\begin{figure}
  \begin{center}
    \includegraphics[width=0.3\linewidth]{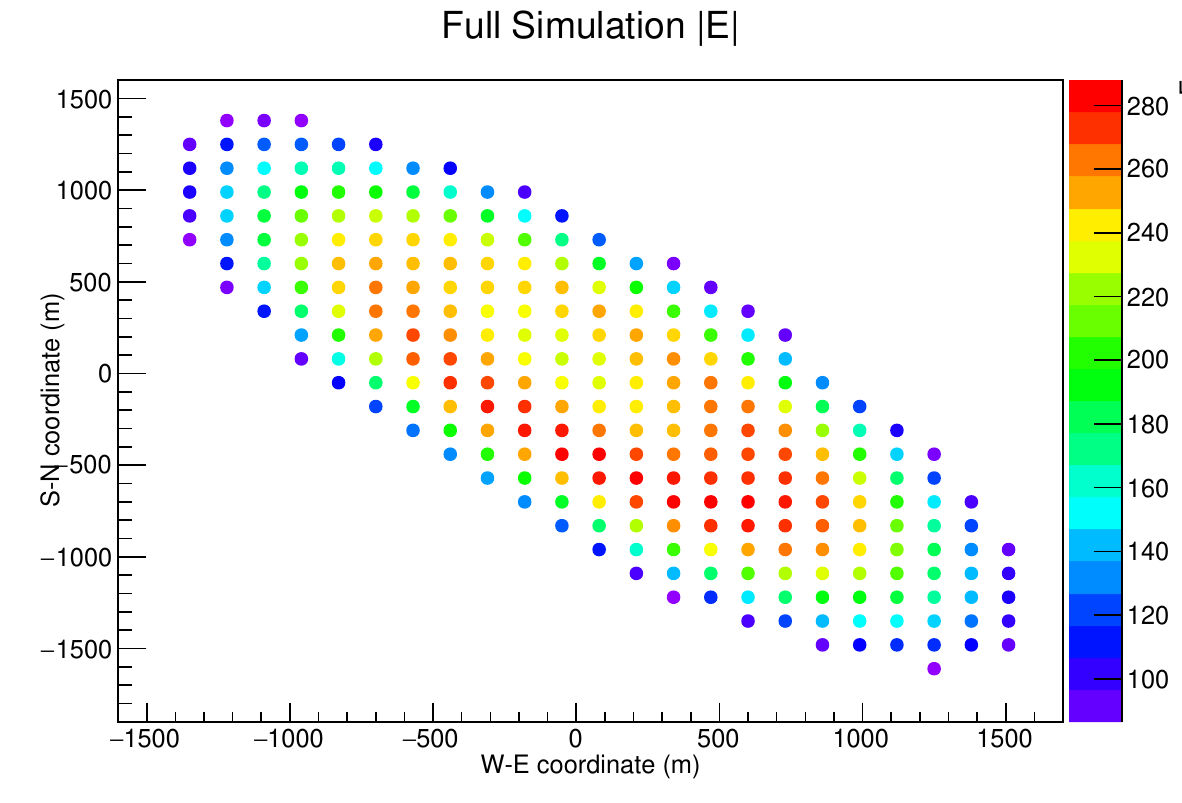}\includegraphics[width=0.3\linewidth]{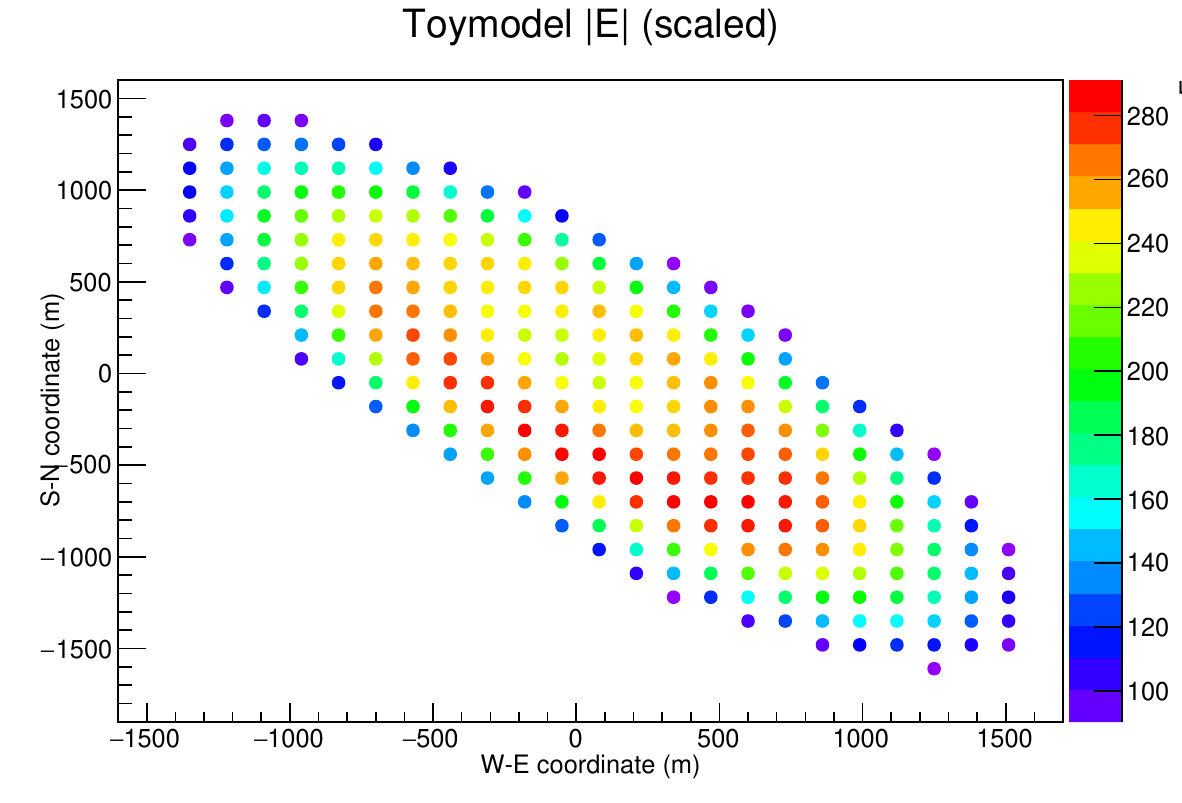}\includegraphics[width=0.3\linewidth]{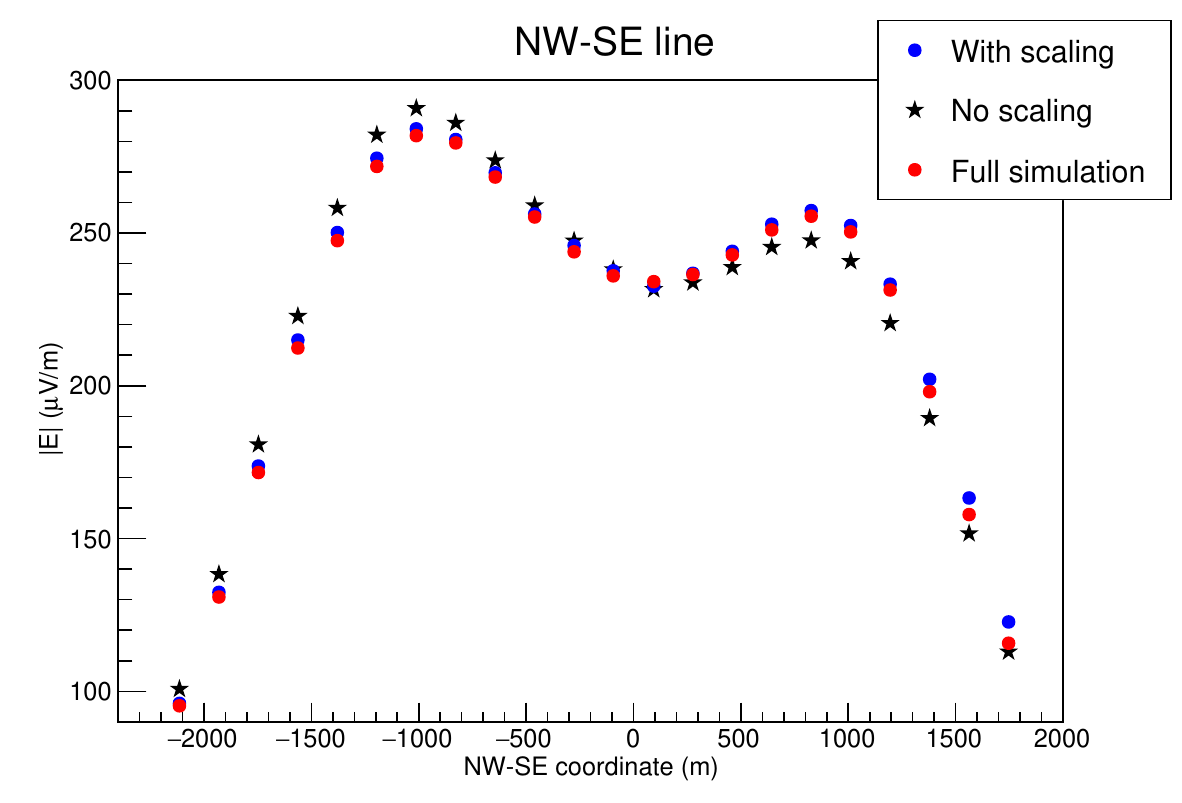}
  \end{center}
  \vspace{-0.6cm}
  \caption{Comparison between full simulations and the superposition model for a 70$^\circ$ proton shower of energy 1 EeV. Left: Full ZHAireS simulation. Middle: RDSim emission model for the same antennas as the full simulation. On both plots the color scale represents the peak electric field in $\mu V/m$. Right: Comparison of the peak electric field along a line in the direction of the major axis of the elliptical footprint for the full simulation (Red), the full emission toymodel (Blue) and the original version of the toymodel~\cite{toymodel}, that did not include early-late effects (stars). The maximum difference in this example is 8\% on just a couple of antennas at the very edge of the footprint, where the signal is very low. A more characteristic maximum error for the vast majority of antennas, even at the edges, is around 4\%. A similar comparison, but for a 80$^\circ$ shower can be seen in \cite{RDSimARENA}.  }
  \label{fig:fullsimcomparison}
\end{figure}

To make it possible to use a single ZHAireS simulation for multiple arrival directions, RDSim's emission model can be rotated to any desired azimuth angle. This rotation takes into account all relevant parameters, such as the change of the angle $\alpha$ between the shower axis and the geomagnetic field and the changes in the distance between the simulated antennas on the reference line and the shower maximum. The maximum errors introduced by this rotation are very small, around 2\%. For more details see \cite{RDSimARENA}. We have also compared the results of the superposition model with full ZHAireS simulations and find a very good agreement between the two, as can be seen on Fig.~\ref{fig:fullsimcomparison}, where we show a comparison between the electric field obtained with full simulations and the superposition model for a 70$^\circ$ 1 EeV proton shower.

\section{Detector response and structure}
\label{sec:detection}

We model the characteristics of the array and its antennas in a very simple way. RDSim can be used with any arbitrary array with antenna positions on a horizontal plane at ground level. All antennas in the array are assumed to be identical and a simple threshold in electric filed amplitude is then set to trigger the antennas.  We can also consider the effect of the beam pattern of the antennas in the trigger. To do this we simply multiply the electric filed obtained from the radio emission model by the beam pattern at the arrival zenith angle. This effective electric field is then used to evaluate the antenna trigger. For an array-level trigger we simply use a settable minimum number of antenna-level triggers required to trigger the whole event. 
A more detailed description can be seen in~\cite{RDSimARENA}.

\begin{figure}
  \begin{center}
    \includegraphics[width=0.4\linewidth]{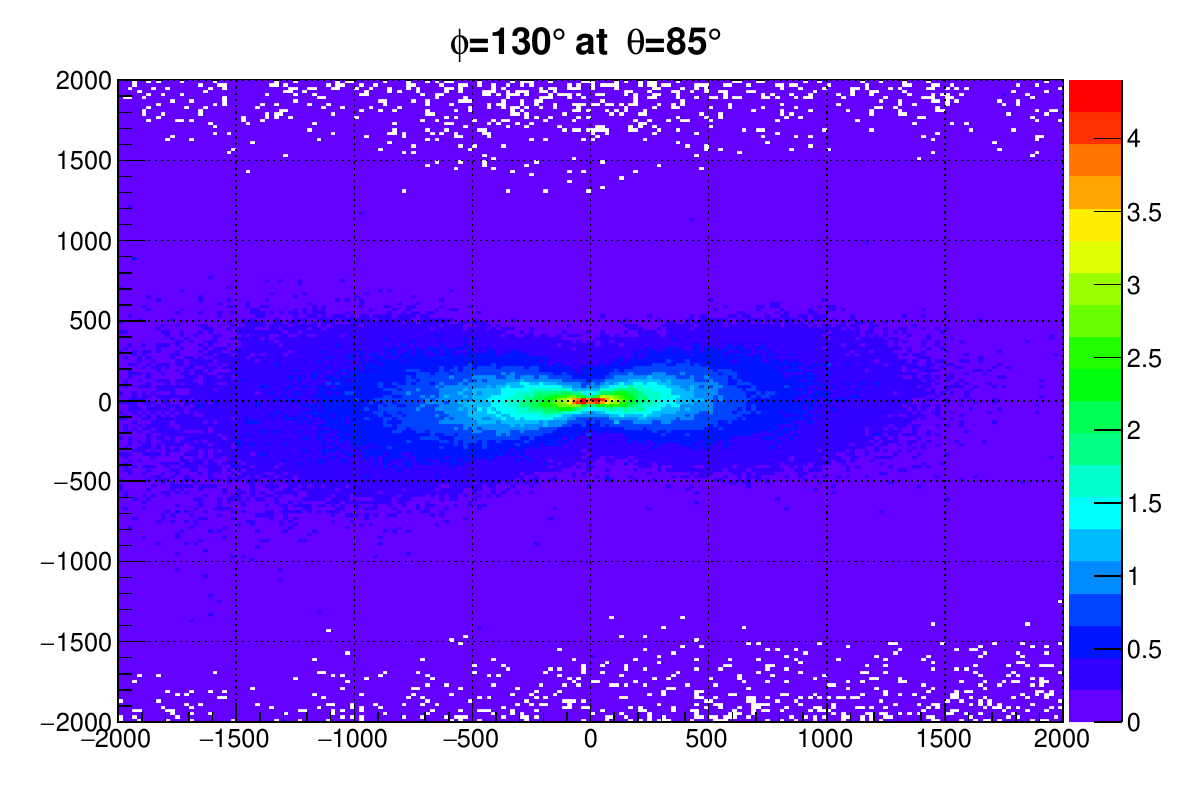}\includegraphics[width=0.4\linewidth]{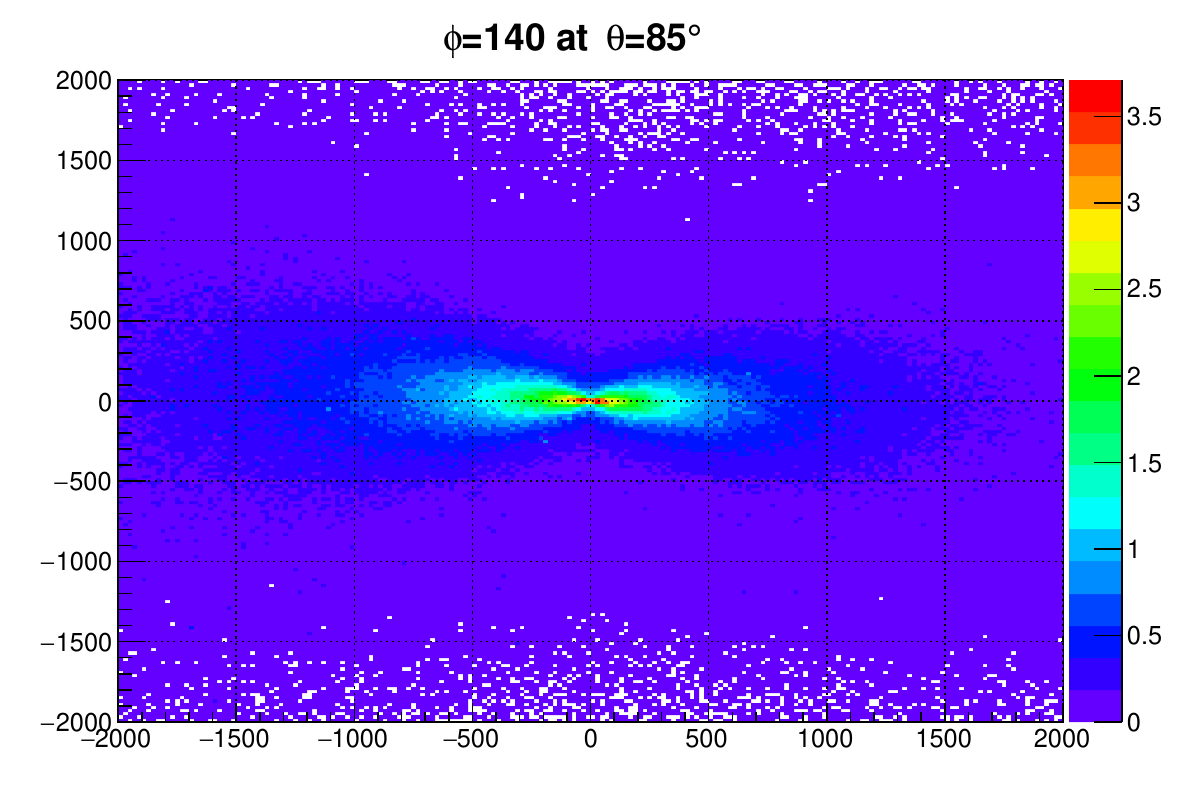}
    \includegraphics[width=0.4\linewidth]{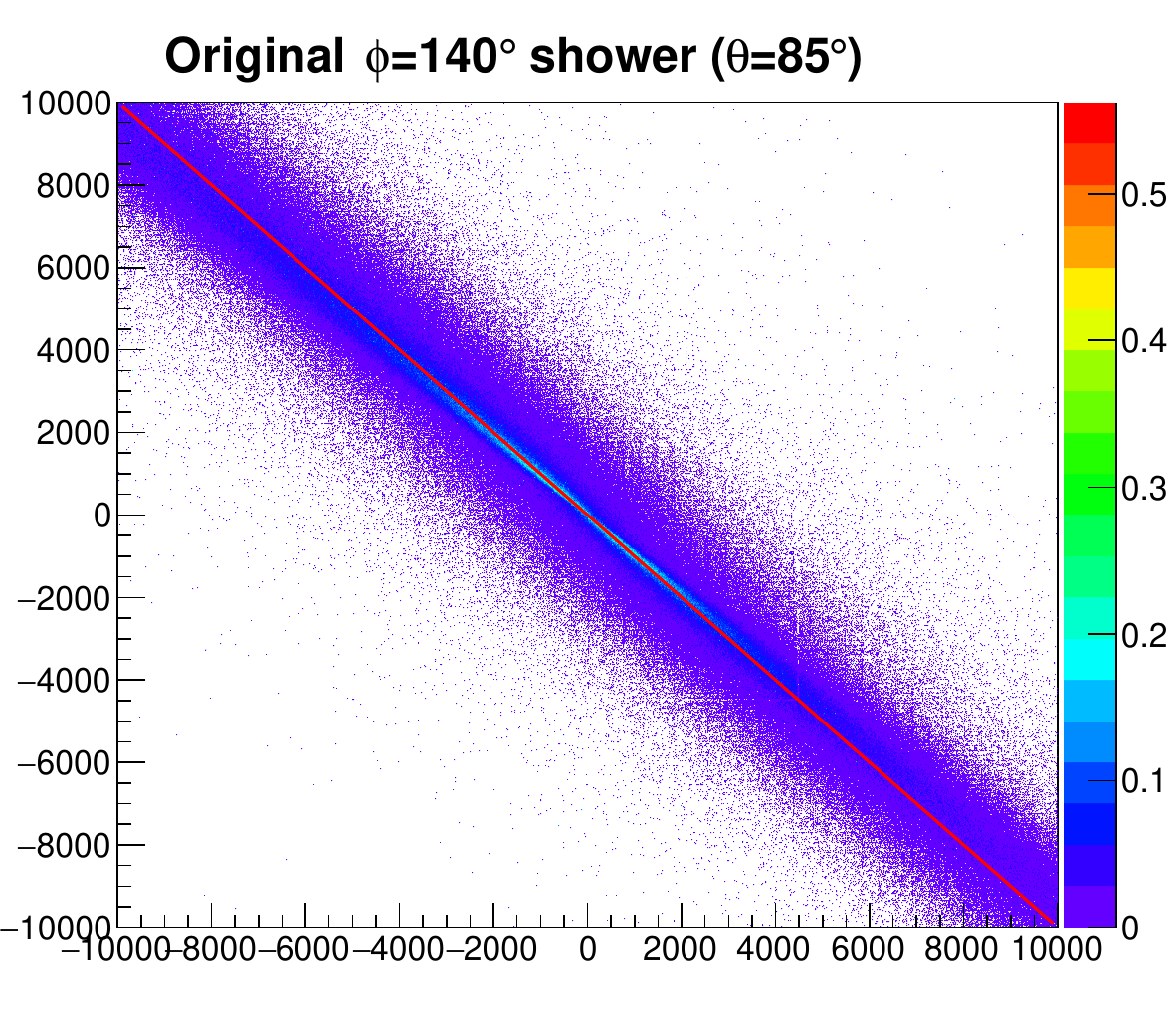}\includegraphics[width=0.4\linewidth]{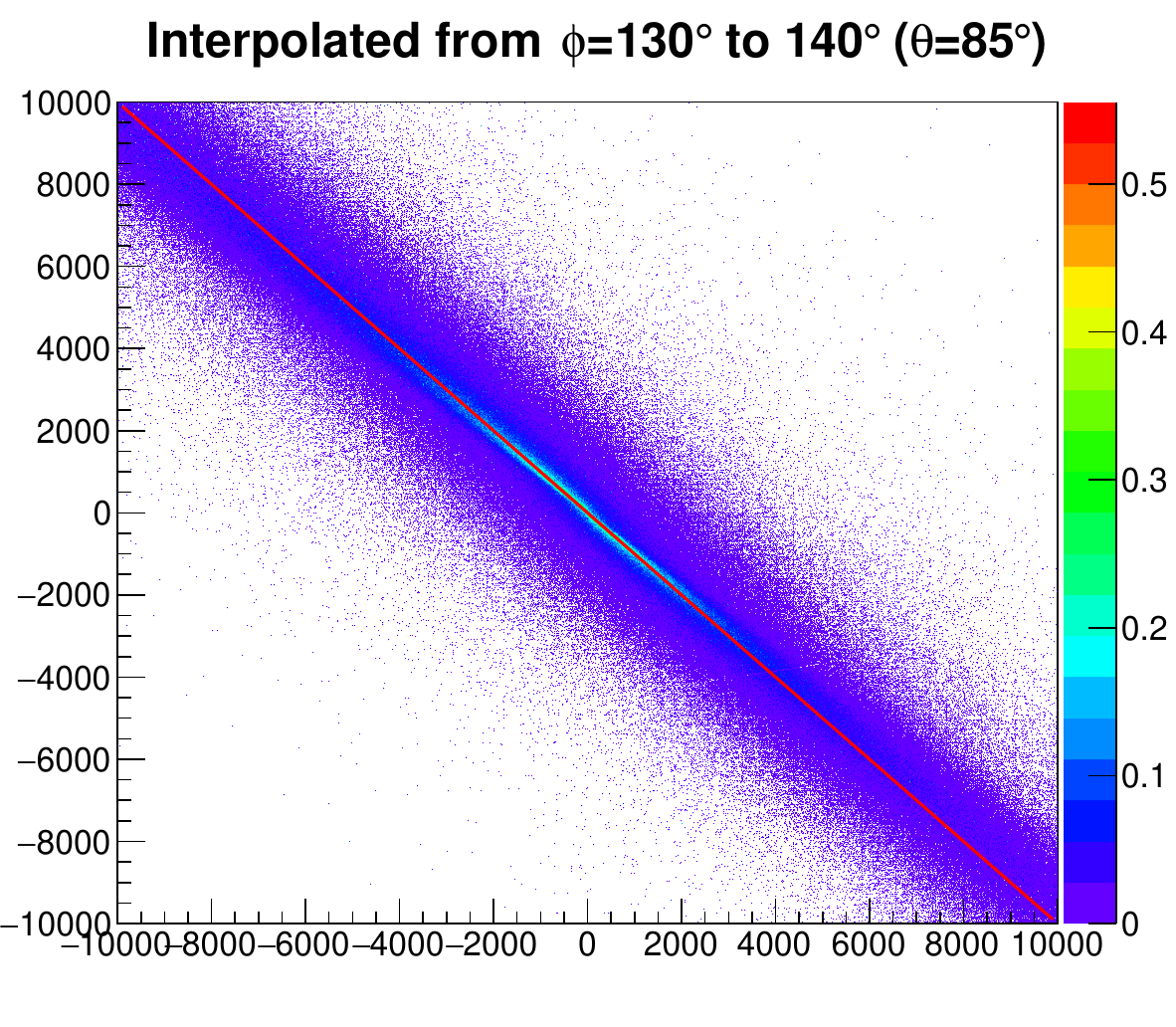}
    \includegraphics[width=0.4\linewidth]{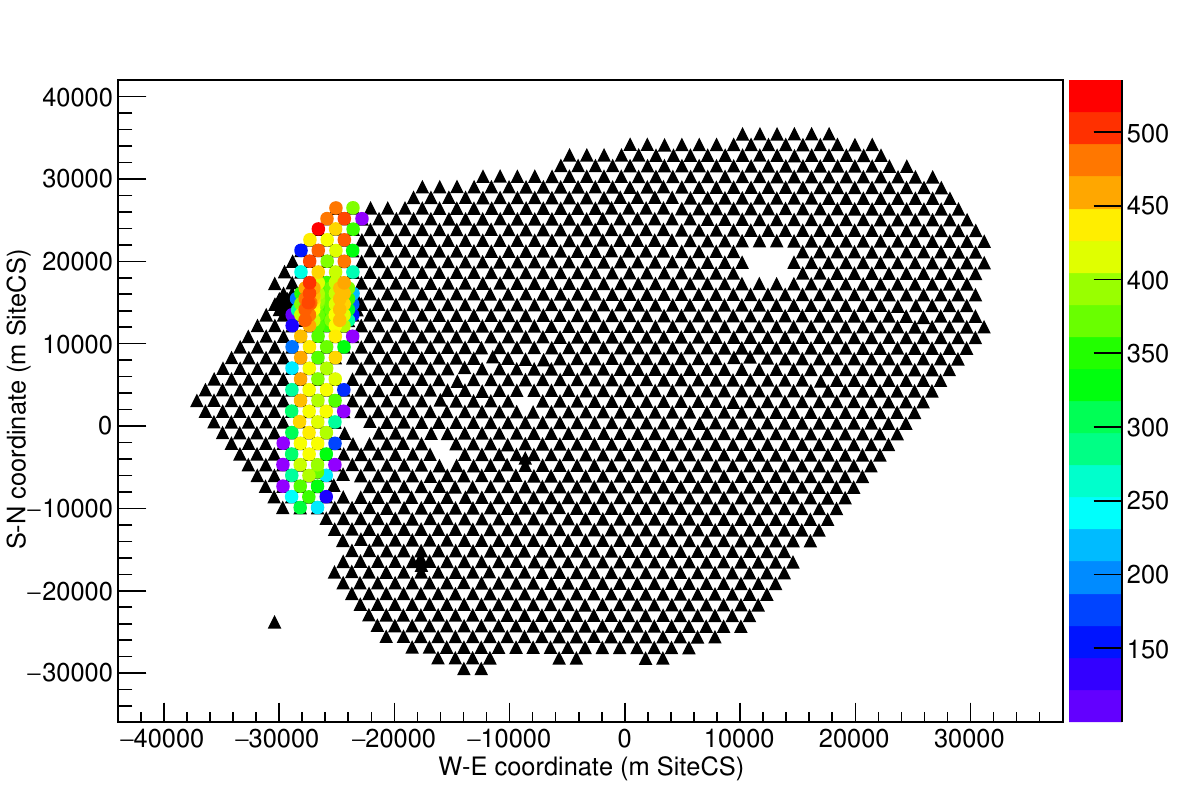}\includegraphics[width=0.4\linewidth]{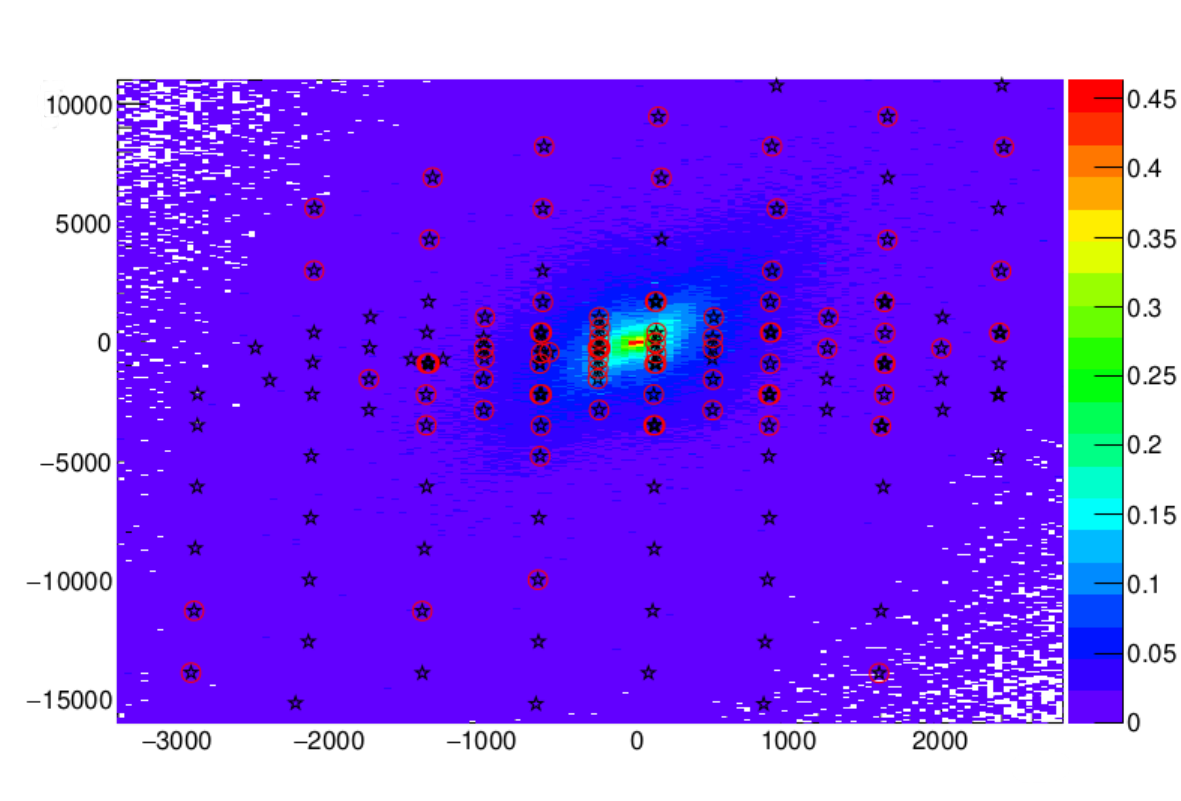}
  \end{center}

  \vspace{-0.6 cm}
  
  \caption{Top: Muon densities on the perpendicular plane of an 85$^\circ$ zenith angle shower with azimuth 130$^\circ$ (left) and azimuth 140$^\circ$ (right). The x and y axes are the coordinates in m on the perpendicular plane and the color scale represents the number of muons per $m^2$. Middle: Muon densities at ground level for an unrotated shower with azimuth 140$^\circ$ (left) and a 130$^\circ$ azimuth shower rotated to 140$^\circ$ (right). Bottom left: Example event at Auger-RD~\cite{AugerPrime}. The color scale is the effective peak electric field amplitude in $\mu V/m$. Bottom right: particle simulation of the same event, where the color scale represents the ground level muon density in muons/$m^2$ used to sample the number of muons crossing each detector. Station positions are shown as stars and triggered detectors are marked with a red circle. Coordinates are in meters w.r.t. the core position. In this example the threshold for the trigger was set to a single muon crossing the detector. }
  \label{fig:MuMapRotation}
\end{figure}

Some detectors also require ground particles to trigger. To accommodate for this, we have implemented a simple particle trigger. Currently, RDSim only takes into account high energy muons by using a simple model to estimate the muon density at ground level. This model uses as input AIRES simulations of ground particles at a low thinning level. Similarly to the radio emission model, we have also implemented a rotation of the muon model to make it possible to use a single AIRES simulation for many arrival directions. To do this we project the ground muons obtained from the simulation into the perpendicular plane of the simulated arrival direction $(\theta,\phi)$. Small variations in the arrival azimuth leads to only very small variations in the angle $\alpha$ between the shower axis and the geomagnetic field. This leads to only small variations in the Lorenz force intensity and direction and thus only small variations in the perpendicular plane muon map. This can be visualized on the top panel of Fig.~\ref{fig:MuMapRotation}, where we show the muon density on the perpendicular plane for a simulation at $\phi=130^\circ$ (left) and one at $\phi=140^\circ$ (right), for the same 85$^\circ$ zenith angle.  Based on this approximation, the muon model assumes that the simulated muon map on the perpendicular plane can be used to produce muon maps at ground level for other arrival directions, provided the changes in the azimuth angle are small. In order to rotate a ground muon simulation performed for an azimuth angle $\phi$ to a different azimuth $\phi'$ we simply use the same original perpendicular plane map obtained for $(\theta,\phi)$, but project it back to the ground using the new arrival direction $(\theta,\phi')$. The middle panel of Fig.~\ref{fig:MuMapRotation} shows the muon densities at ground level for an unrotated shower with azimuth 140$^\circ$ (left) and a 130$^\circ$ azimuth shower rotated to 140$^\circ$ (right). The default behavior for this procedure is to allow a maximum difference in azimuth of 10$^\circ$ ($|\phi-\phi'|<10^\circ$) for the rotation. This means that performing AIRES ground muon simulations only every 20$^\circ$ in azimuth allows the rotation of the maps to any desired azimuth.

Once the muon density map at ground level is obtained for a given event, we estimate the number of muons crossing each particle detector. To take into account the particle detector shape, we calculate an effective area $A_{\mbox{eff}}(\theta)$ as a function of the shower zenith angle, which represents the area on the ground that is shadowed by the detector for a given arrival direction. As an example, a circular Auger-like tank with radio $r$ and filled with water to a height $h$ would have an effective area $A_{\mbox{eff}}(\theta)=\pi r^2 + 2rh\tan{\theta}$, while for a horizontal scintillator on the ground the effective area is just its geometrical area. We then sample the number of muons crossing the detector from a Poisson distribution with a rate parameter $\lambda = A_{\text{eff}}(\theta)\rho_{\mu}$, where $\rho_{\mu}$ is the muon density at the location of the detector. To evaluate the particle detector trigger state we use a simple settable trigger threshold on the number of muons that crosses the detector. This particle simulation is performed after the main radio-only detection part of the simulation finishes and is calculated only for those events and stations that triggered in the radio-only part. This approach simplifies the main RDSim run and keeps its high speed intact. An example simulated event at Auger-RD~\cite{AugerPrime} can be seen on the bottom panel of Fig.~\ref {fig:MuMapRotation}. On the left we show the result of the radio-only part of the simulation, while on the right we show the simple particle trigger simulation for the same event.

RDSim was built around speed and is implemented in C++ with just a few key classes. The main run is controlled by an input file which contains parameters such as arrival direction, energy and core position ranges along with the settable trigger thresholds. It also contains links to the previously produced instances of the radio emission model along with ranges for their validity. This main input file also sets the secondary input files, such as the optional antenna pattern file and, in the case of events induced by tau decays, the file with the parametrizations for the tau propagation (see section~\ref{sec:neutrinos}). During the run, and for each event, RDSim samples an energy, core position and arrival direction.  In the case of neutrino events it also samples the interaction or decay depth where the shower starts. Once the parameters of the event are sampled, RDSim searches for all radio emission toymodel instances that can be used for that particular event and chooses one randomly. The chosen toymodel is rotated and scaled to match the parameters of the event and then used to calculate the electric field at each antenna. The (optional) beam pattern is then applied to the calculated fields and the trigger conditions are checked. When the main run ends, the complete information of each event is saved in a compressed ROOT file containing, among other parameters, the event number, arrival direction, core position, energy, number of triggered stations, the details of the specific instance of the radio emission model that was used and in the case of neutrino events also the interaction or decay depth. For events that triggered, the peak electric field of each triggered antenna is also saved. More details can be seen in \cite{RDSimARENA}.



\section{Neutrino events in RDSim}
\label{sec:neutrinos}

RDSim can handle all neutrino interaction channels, but in order to simulate neutrino events some extra parameters are needed.  In the case of showers initiated by CC and NC interactions, we use an extensive library of HERWIG \cite{herwig} simulations of neutrino interactions. The products of the simulation are injected into ZHAireS to calculate the radio emission of neutrino induced events. Since the neutrino cross-section is very small for all energies, we assume the point of neutrino interaction to be equally distributed in atmospheric depth $X$. So, in order to sample the point in the atmosphere where the neutrino interacts and the shower starts, we simply divide the atmosphere in slices of equal thickness $\Delta X$ in atmospheric depth, centered at a various interaction depths $X_{\text{int}}$. Instances of the superposition emission model are then created for each slice from ZHAireS simulations. RDSim then chooses one of the slices at random, with equal probability, and one of the corresponding instances of the emission model at that particular $X_{\text{int}}$ to simulate the radio emission of the shower.

In case of showers induced by tau-lepton decays, the procedure is similar but TAUOLA \cite{tauola} simulations of tau decays are used instead of HERWIG as input for the air shower simulations. In order to sample the position where the shower starts, i.e. the point where the tau decays, one has to take into account the propagation of the tau from its creation, where the $\nu_{\tau}$ interacts, until its decay. The $\nu_{\tau}$ interaction depth $X_{0}$ in the atmosphere is sampled as before by choosing a random atmospheric $\Delta X$ slice. We then propagate the tau, disregarding its energy losses in air, to obtain the distribution of the decay depths  $X_{\mbox{decay}}$, where the showers start. This propagation is based on the probability $dP(E_{\tau})$ of tau decay per meter. In order to maintain RDSim's high speed in this type of event, these propagation simulations are done prior to the main run in order to obtain, for each zenith angle, the probability of the tau decaying before reaching the ground along with a parametrization of the decay depths  $X_{\mbox{decay}}$ for those that decay above ground. The latter also takes into account the position of the $\nu_{\tau}$ interaction. During the main run we sample the probability of the tau decaying before reaching the ground. If it decays above ground we sample the previously produced $X_{\mbox{decay}}$ parametrization and simulate a shower at that position, otherwise the event is instantly marked as not triggered, since no shower is produced.



\section{Discussion}
\label{sec:discussion}

\begin{figure}
  \begin{center}

    \vspace{-0.6cm}
    \includegraphics[width=0.352\linewidth]{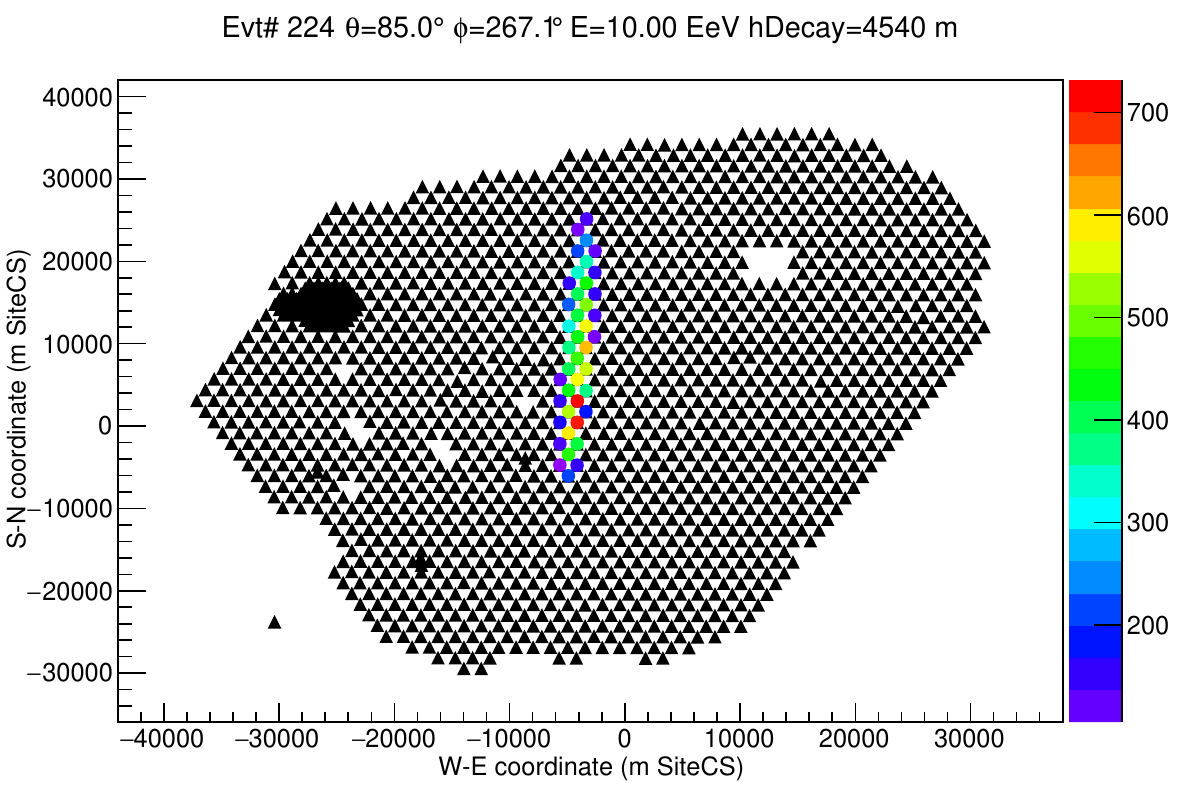}\includegraphics[width=0.352\linewidth]{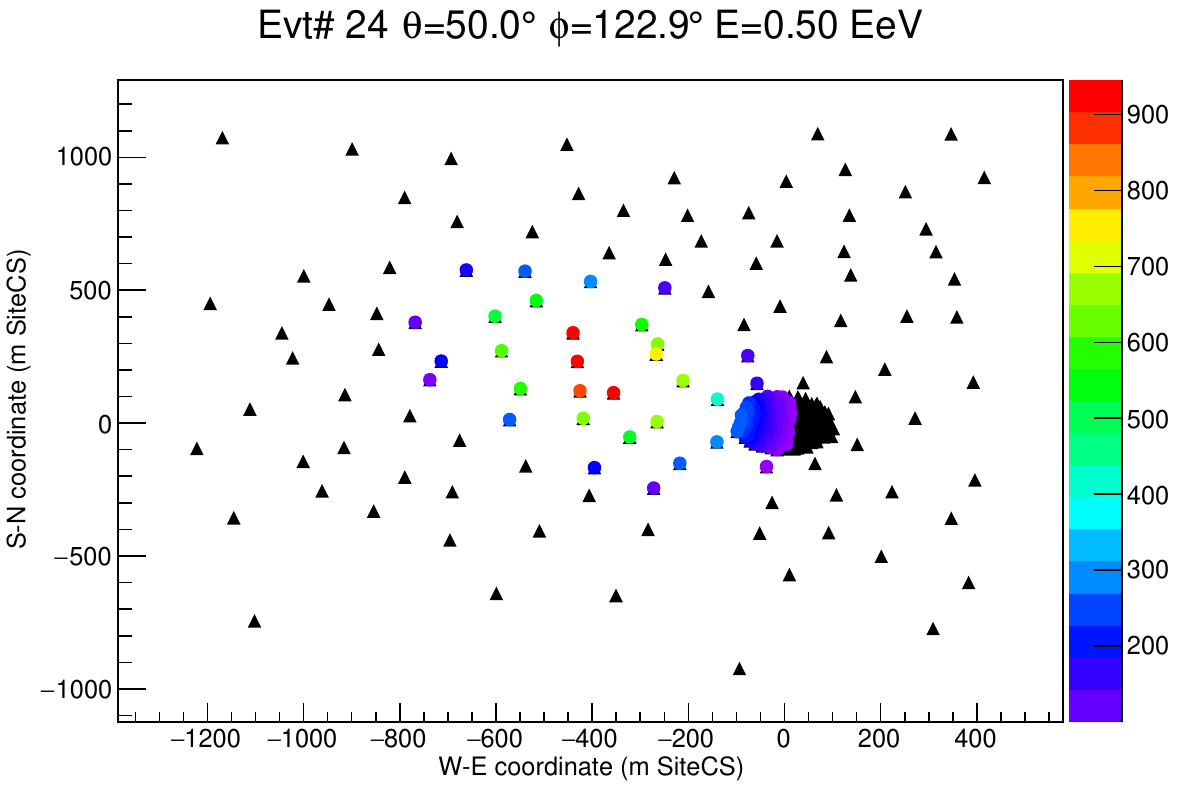}\includegraphics[width=0.28\linewidth]{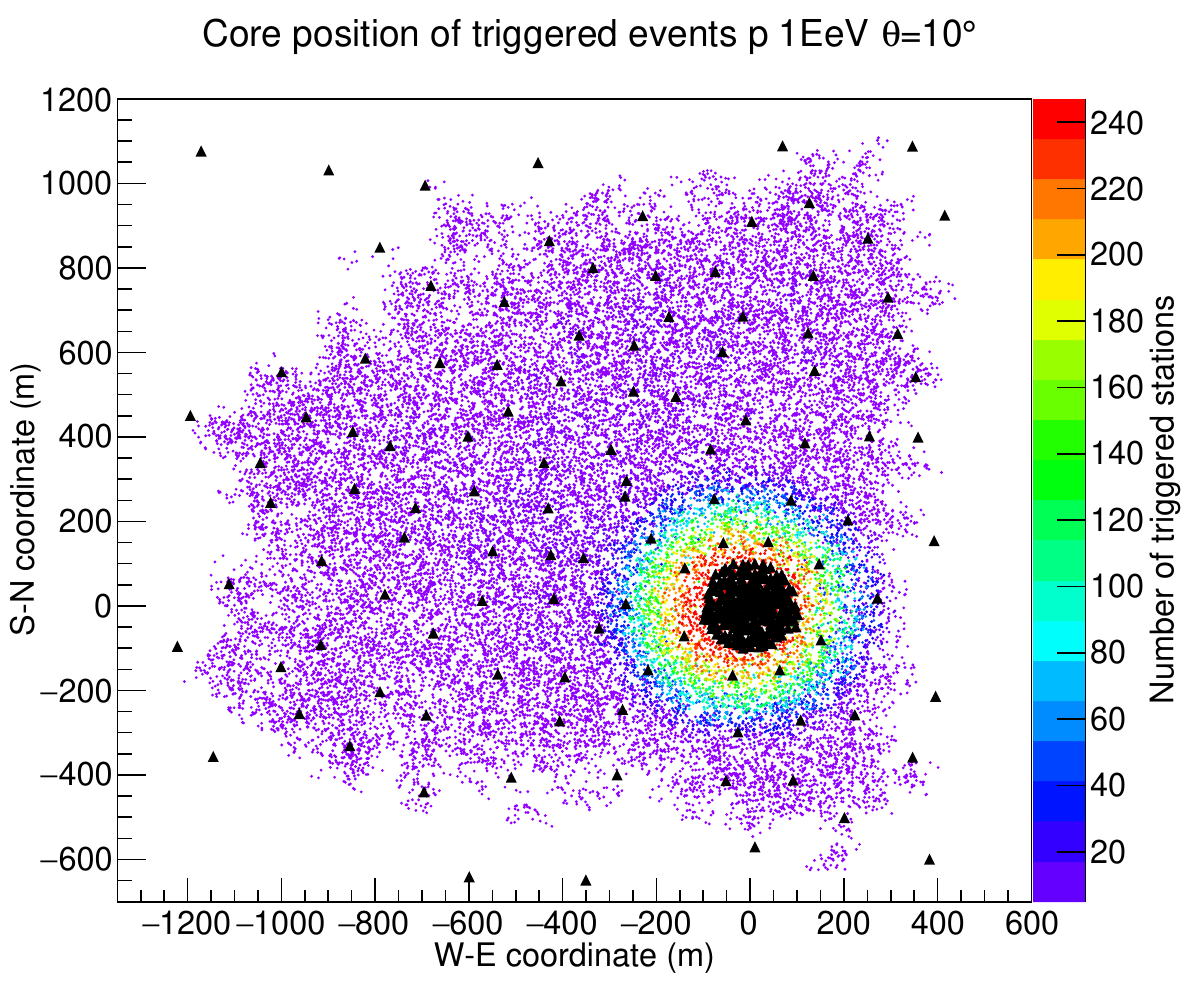}
  \end{center}
  \vspace{-0.6cm}
  \caption{Left: Example event induced by an 85$^\circ$ tau decay occurring $\sim4.5$ km above the AUGER-RD array. Middle: Example of a 50$^\circ$ 0.5 EeV proton event at OVRO-LWA. In both plots the color scale represents the peak electric field in $\mu V/m$. Right: Number of triggered stations as a function of core position for 10$^\circ$ 1 EeV proton showers at OVRO-LWA. One can see that OVRO-LWA cannot trigger for 10$^\circ$ showers landing on the edges of the array, specially towards the NW and S, due to the low antenna density in those regions. But as the core position closes the central part of the array, the number of triggered stations rises very fast due to the large antenna density on this region. }
  \label{fig:examples}
\end{figure}

The RDSim framework is very fast and flexible. It can simulate the radio emission of downgoing air showers and its detection by any horizontal ground antenna array. On Fig.~\ref{fig:examples}, we present two example events, one simulated using the AUGER-RD~\cite{AugerPrime} array (left) and the other using OVRO-LWA~\cite{OVROARENA} (right), two very distinct detectors. RDSim was built with speed in mind and uses several simple, yet still very accurate, toymodel-like approaches. Once it is setup, it is able to simulate millions of events in just a few minutes. Its speed makes it possible to have enough statistics to study in detail the detection probability of every class of event, including events with a very low probability of detection. It can thus be used to shed light into the effect the array characteristics have in its detection capabilities and is specially suited to be used as a fast aperture calculator. It also allows to study geometrical effects in detail, such as those that arise due to border effects or array asymmetries. As a very simple example of this type of study, we show on the right panel of Fig.~\ref{fig:examples} the number of triggered stations as a function of core position for 10$^\circ$ 1 EeV proton showers at OVRO-LWA.   

Another use of RDSim is to optimize the parameters of dedicated full simulation libraries in the event more detailed modeling of the emission and detector response are desired. Since it can calculate the detection probability of every class of event with great precision, it can be used to set the full simulation parameters that would cover the whole phase-space of detectable events, not only estimating the total number of full simulations needed but also the relative number needed for each class of event. This is specially relevant for neutrino events, since they have extra variables if compared to normal showers, such as the variable interaction or decay depth. This makes the phase-space for this type of event much larger. In this case, unoptimized full simulation libraries would either not cover the whole phase-space or be computationally unfeasible due to the shear number of unoptimized simulations needed to cover it fully.

We are in the process of comparing RDSim simulated events with full simulations of both the radio emission and detector response. Our preliminary results shows a very good agreement between RDSim and full simulations, despite the huge decrease in computing time. We are also extending RDSim to simulate mountain events. These are events induced by the decay of tau-leptons that are produced by $\nu_\tau$ interactions inside mountains around the detector. This will be accomplished by using topographical maps to calculate the amount of rock traversed and the distance to the closest rock face for any given core position and arrival direction. This would be all that is needed not only to simulate such events, but also to convolute RDSim results with the probability of a given tau to be produced and exit the mountain on an event-by-event basis.

\end{document}